\documentclass{article}
\usepackage{ijcai22}

\usepackage{times}
\usepackage{soul}
\usepackage{url}
\usepackage{thm-restate}
\usepackage{thmtools} 

\usepackage[hidelinks]{hyperref}
\usepackage[utf8]{inputenc}
\usepackage[small]{caption}
\usepackage{graphicx}
\usepackage{amsmath}
\usepackage{amsthm}
\usepackage{booktabs}
\usepackage{algorithm}
\usepackage{algorithmic}
\usepackage{mleftright}
\usepackage{bm}
\usepackage{xcolor}
\usepackage{amsmath}
\usepackage{amsthm}
\usepackage{amsfonts}
\usepackage{bbm}
\usepackage{mathtools}
\usepackage{comment}

\usepackage[normalem]{ulem}

\usepackage{pgfplots}
\pgfplotsset{compat=1.15}
\usepackage{mathrsfs}
\usetikzlibrary{arrows}
\newtheorem{example}{Example}
\newtheorem{theorem}{Theorem}

\newtheorem{proposition}{Proposition}

\newtheorem{assumption}{Assumption}

\DeclareMathOperator*{\argmax}{arg\,max}

\DeclareMathOperator*{\GSP}{\textnormal{GSP}}
\DeclareMathOperator*{\VCG}{\textnormal{VCG}}
\DeclareMathOperator*{\w}{\textnormal{w}}

\title{The Power of Media Agencies in Ad Auctions:\\ Improving Utility through Coordinated Bidding}

\author{
}

\author{
Giulia Romano
\and
Matteo Castiglioni \and
Alberto Marchesi \and
Nicola Gatti
\affiliations
	Politecnico di Milano, Piazza Leonardo da Vinci 32, I-20133, Milan, Italy 
\emails
\{giulia.romano, matteo.castiglioni, alberto.marchesi, nicola.gatti\}@polimi.it
}

\begin{document}

\maketitle

\begin{abstract}

The increasing competition in digital advertising induced a proliferation of \emph{media agencies} playing the role of intermediaries between \emph{advertisers} and \emph{platforms} selling ad slots. 
 When a group of competing advertisers is managed by a common agency, many forms of collusion, such as bid rigging, can be implemented by coordinating bidding strategies, dramatically increasing advertisers' value.
 We study the problem of finding \emph{bids} and \emph{monetary transfers} maximizing the utility of a group of colluders, under GSP and VCG mechanisms.
 First, we introduce an abstract bid optimization problem---called \emph{weighted utility problem} (WUP)---, which is useful in proving our results.
 We show that the utilities of bidding strategies are related to the length of paths in a directed acyclic weighted graph, whose structure and weights depend on the mechanism under study.
 This allows us to solve WUP
 in polynomial time by finding a shortest path of the graph.
 Next, we switch to our original problem, focusing on two settings that differ for the \emph{incentives} they allow for.
 Incentive constraints ensure that colluders do \emph{not} leave the agency, and they can be enforced by implementing monetary transfers between the agency and the advertisers.
 %
 In particular, we study the \emph{arbitrary transfers} setting, where any kind of monetary transfer to and from the advertisers is allowed, and the more realistic \emph{limited liability} setting, in which no advertiser can be paid by the agency.
 In the former,
 we cast the problem as  
 a WUP instance and solve it by our graph-based algorithm, while, in the latter,
 we formulate it as a 
 linear program with exponentially-many variables 
 efficiently solvable by applying the ellipsoid algorithm to its dual.
 This requires to solve a suitable separation problem in polynomial time, which can be done by reducing it to 
 a WUP instance.

\end{abstract}

\section{Introduction}

Over the last years, digital advertising has been one of the main drivers of the growth of world market.
Remarkably, the vast majority of the companies employ digital tools to advertise their products or services, and the worldwide annual spent in digital advertising reached about 150 billion USD in 2020~\cite{iab2020iab}. 
Furthermore, most of the economic reports forecast that \emph{artificial intelligence} (AI) will fuel an increase of market value by almost 100\% over the next decade~\cite{McKinsey2018}. 
Indeed, AI tools have become widespread throughout digital advertising, enabling new opportunities that were not available before. Notable examples are in, \emph{e.g.}, auction design~\cite{Bachrach2014Optimising}, budget allocation~\cite{Nuara2018Combinatorial}, bidding opimization~\cite{He2013Game}, and multi-channel advertising~\cite{Nuara2019Dealing}.

Recent years have witnessed a proliferation of \emph{media agencies}, which claim to play the role of intermediaries between \emph{advertisers} and \emph{platforms} selling ad slots.
This has been driven by the increasing complexity of digital advertising---due to, \emph{e.g.}, a huge amount of available data and of parameters to be set on advertising platforms---and the rising competition among a growing number of advertisers.
%
%
When a group of competing advertisers is managed by a common agency, it frequently happens that the agency has to place bids on behalf of different advertisers participating in the same ad auction.
This dramatically changes the strategic interaction underlying ad auctions, since agencies can coordinate advertisers' bidding strategies by implementing many forms of collusion---such as, \emph{e.g.}, bid rigging--, and, thus, they can increase advertisers' value.
%
%
In particular, simple examples show that colluding in ad auctions can reward the colluders with a utility that is arbitrarily larger than what they would get without doing that.
Moreover, a recent empirical study on real-world data by one of the major US agencies (Aegis-Dentsu-Merkle) shows that collusion is pervasive and leads to a significant reduction in the average cost-per-click \cite{RePEc:net:wpaper:1708}.

\paragraph{Original Contributions.}
We study the computational problem faced by a media agency that has to coordinate the bidding strategies of a group of colluders, under GSP and VCG mechanisms (details on the adoption of such mechanisms in ad auctions are provided in~\cite{10.1257/aer.104.5.442}).
We assume that the media agency knows the private valuations of the colluders (\emph{i.e.}, how much they value a click on their ad), and that it decides colluders' bids on their behalf. 
Moreover, the media agency is in charge of paying the auction mechanism for a click on an allocated ad, and at the same time it requires \emph{monetary transfers} to and from the colluders.
These are necessary to enforce some \emph{incentive constraints} ensuring that the colluders do \emph{not} leave the agency and participate in the ad auction individually. 
In this paper, we study two settings that differ on the kind of monetary transfers that they allow for: the \emph{arbitrary transfers} setting, where any kind of monetary transfer to and from the advertisers is allowed, and the more realistic \emph{limited liability} setting, in which no advertiser can be paid by the media agency.
Finally, we assume that the bids of the advertisers external to the media agency are drawn according to some probability distribution. 
As a first result, we introduce an abstract bid optimization problem---called \emph{weighted utility problem} (WUP)---, which works for any finite set of possible bid values and is useful in proving our main results in the rest of the paper.
In order to solve such a problem, we first show that the utilities of bidding strategies are related to the length of paths in a directed acyclic weighted graph, whose structure and weights depend on the mechanism under study (either GSP or VCG mechanism).
This allows us to solve  
WUP instances in polynomial time by finding a shortest path of the graph.
Next, we switch the attention to the original media agency problem, starting from the arbitrary transfers setting.
A major challenge is dealing with a potentially continuous set of possible bids.
%
Notably, we show that it is possible to reduce the attention to a finite set of bidding strategies, only incurring in a small additive loss in the value of the obtained solution and relaxing the incentive constraints by a small additive amount.
The set is built by recursively splitting the interval of possible bid values, until one gets sub-intervals such that the probability that an external bid is in a given sub-interval is sufficiently small. 
Then, the resulting sub-intervals are used to define the desired finite set of bids.
In conclusion, we cast the problem as a
WUP instance
and solve it by our graph-based algorithm in polynomial time.
This gives a bi-criteria FPTAS for the original problem, since the approximation is in terms of both objective value and incentive constraints.
Finally, we study the limited liability setting.
In this case, we leverage the same finite set of bidding strategies defined for the arbitrary transfers setting in order to formulate the problem as a \emph{linear program} (LP) with exponentially-many variables and polynomially-many constraints.
Since we use only a finite set of bids, we need to relax the incentive constraints by an arbitrary small amount to guarantee the existence of a feasible solution.
%
We solve the LP in polynomial time by applying the ellipsoid algorithm~\cite{grotschel1981ellipsoid} to its dual, which features polynomially-many variables and exponentially-many constraints.
This requires solving a suitable separation problem in polynomial time, which can be done by reducing it to 
WUP instance. As in the arbitrary transfer setting, the resulting algorithm is a bi-criteria FPTAS.

\paragraph{Related Works.}
While a longstanding literature investigates the role of mediators in ad auctions---see, \emph{e.g.}, the seminal works by \citeauthor{Vorobeychik2008}~[\citeyear{Vorobeychik2008}] and \citeauthor{Ashlagi2009MediatorsIP}~[\citeyear{Ashlagi2009MediatorsIP}]---, collusion is currently emerging as one of the central problems in advertising as the adoption of AI algorithms can concretely support an agency to find the best collusive behaviors~\cite{OECD}. 
Motivated by the recent study by \citeauthor{RePEc:net:wpaper:1708}~[\citeyear{RePEc:net:wpaper:1708}], some works provide theoretical contributions to assess how collusion can be conducted by an agency.
\citeauthor{RePEc:inm:ormnsc:v:66:y:2020:i:10:p:4433-4454}~[\citeyear{RePEc:inm:ormnsc:v:66:y:2020:i:10:p:4433-4454}] study a setting in which there is no monetary transfer between the agency and bidders and provide equilibrium conditions. 
They show that, in simple settings, GSP is more inefficient than VCG both in terms of efficiency and revenue. \citeauthor{lorenzon}~[\citeyear{lorenzon}] studies a setting with two slots and three bidders which are all controlled by an agency in a GSP auction. Furthermore, a monetary transfer is possible. The author shows collusive stable behaviors in which the redistribution is uniform over the three bidders.

\section{Preliminaries}

We study the problem of coordinated bidding faced by a media agency in ad auctions, with both GSP and VCG payments.
In this setting, there is a set $N_c \coloneqq \{1,\ldots,n_c\}$ of \emph{advertisers} (or \emph{agents}), whose advertising campaigns are managed by a common media agency. 
Moreover, there is another set $N_e \coloneqq \{1,\ldots,n_e\}$ of advertisers who are \emph{not} part of the agency, but participate to the ad auction individually.
In this work, we refer to the former as \emph{colluders}, while we call the latter \emph{external agents}.
For the ease of presentation, we let $N \coloneqq N_c \cup N_e$, and, with a slight abuse of notation, we sometimes write $N \coloneqq \{1,\ldots,n\}$ by renaming the agents, where $n=n_c+n_e$.
The advertisers compete for displaying their ads on a set $M \coloneqq \{1,...,m\}$ of \emph{slots}, with $m \leq n$.
Each agent $i \in N$ has a \emph{private valuation} $v_i \in [0,1]$ for an advertising slot, which reflects how much they value a click on their ad.
Moreover, each slot $j \in M$ is associated with a \emph{click through rate} parameter $\lambda_j \in[0,1]$, encoding the probability with which the slot is clicked by a user.\footnote{For the ease of presentation, we assume that the click through rate only depends on the slot and \emph{not} on the ad being displayed. This dependence can be easily captured by interpreting $v_i$ as an expected value w.r.t.~clicks once the user observed the slot.
} 
Each agent $i \in N$ participates in the ad auction with a \emph{bid} $b_i \in [0,1]$, representing how much they are willing to pay for a click on their ad.
We denote by $b = (b_1, \ldots , b_{n})$ the bid profile made by all the agents' bids.
We also let $b^c = (b^c_1, \ldots , b^c_{n_c})$ be the profile of colluders' bids (also called \emph{bidding strategy}), while $b^e = (b^e_1, \ldots , b^e_{n_e})$ is the profile of external agents' bids.
For the ease of notation, we sometimes write $b = (b^c,b^e)$ to denote the profile made by all the bids in $b^c$ and $b^e$.
Finally, the media agency knows the valuations $v_i$ of all the colluders $i \in N_c$, and it decides the bid profile $b_i^c$ on their behalf.
Additionally, the media agency defines a monetary \emph{transfer} $q_i \in [-1,1]$ for each colluder $i \in N_c$.
We adopt the convention that, if $q_i > 0$, then the transfer is from the agent to the agency, while, if $q_i < 0$, then it is the other way around.\footnote{Notice that there are some scenarios in which it is in the interest of the media agency to pay a colluder in order to ensure that they stay in the agency; see Example~\ref{ex_agencycoord_at} in the Supplementary Material.}

In this work, we assume w.l.o.g. that the bids in a profile $b = (b_1, \ldots , b_{n})$ are ordered by decreasing value of $b_i$, and that the same holds for $b^c$ and $b^e$.
Moreover, w.l.o.g., we assume that the sots are ordered so that $\lambda_1 \geq \ldots \geq \lambda_m$.

The auction goes on as follows.
First, the media agency selects a bidding strategy $b^c = (b^c_1, \ldots , b^c_{n_c})$ and requires a transfer $q_i$ from each colluder $i \in N_c$. 
Then, external agents individually report their bids to the auction mechanism, resulting in a profile $b^e = (b^e_1, \ldots , b^e_{n_e})$, while the media agency reports bids $b^c$ on behalf of the colluders.
Finally, given all the agents' bids $b = (b^c,b^e)$, the mechanism allocates an ad to each slot and defines an \emph{expected payment} $\pi_i(b) \in [0,1]$ for each agent $i \in N$, where the expectation is with respect to the clicks.
The media agency is responsible of paying the mechanism on behalf of the colluders.
%
%

%
Given a bid profile $b = (b_1, \ldots , b_{n})$ and assuming w.l.o.g. that each bidder $i \in [m]$ is assigned to slot $i$, we denote bidder $i$'s \emph{expected revenue} as $r_i(b) \coloneqq \lambda_{i} v_i$, while bidder $i$'s \emph{expected utility} is $u_i(b) \coloneqq r_i(b) - q_i$.\footnote{We denote with $[m]$ the set $\{1, \ldots, m\}$.}
%
%
Instead, the expected utility of the agency is $\sum_{i \in N_c} (q_i - \pi_i(b))$.
We also denote with $U $ the cumulative expected utility of all the colluders and the media agency.
Formally, $U \coloneqq \sum_{i \in N_c} (r_i(b) - \pi_i(b) )$.

Next, we review GSP and VCG mechanisms in ad auctions (see the book by~\citeauthor{nisan2001algorithmic}~[\citeyear{nisan2001algorithmic}] for their general description).
Given a bid profile $b = (b_1, \ldots, b_n)$, both mechanisms orderly assign the first $m$ agents, who are those with the highest bids, to the first $m$ slots, which are those with the highest click through rates.
Moreover, the mechanisms assign the following expected payments.
\begin{itemize}
	\item \emph{GSP mechanism}: $\pi_i^{\GSP}(b) \coloneqq \lambda_i b_{i+1}$ for each agent $i \in [m]$, and $\pi_i^{\GSP}(b) = 0$ for all the other agents.
	\item \emph{VCG mechanism}: $\pi_i^{\VCG}(b) \coloneqq \sum_{j=i+1}^{m+1} b_{j}(\lambda_{j-1} -\lambda_{j} )$ for each agent $i \in [m]$, where we let $\lambda_{m+1} = 0$, and $\pi_i^{\VCG}(b) = 0$ for all the other agents.
\end{itemize}

The VCG payments are such that each agent is charged a payment that is equal to the externalities that they impose on other agents.
This makes the VCG mechanism \emph{truthful}, which means that it is a dominant strategy for each agent to report their true valuation to the mechanism, namely $b_i =   v_i$ for every $i \in N$. 
This is \emph{not} the case for the GSP mechanism.

\section{Problem Formulation}

In this section, we introduce the optimization problem faced by the media agency.
In words, the goal of the media agency is to find a bidding strategy $b^c= (b^c_1, \ldots, b^c_{n_c})$ that coordinates colluders' bids in a way that maximizes the cumulative expected utility $U$, while at the same time guaranteeing that they are incentivized to be part of the media agency.
The rest of the section is devoted to formally defining such a problem.

Before introducing the optimization problem, let us notice that knowing the valuations of all the colluders allows the media agency to improve the cumulative expected utility $U$ with respect to the case in which all the bidders act individually.
%
%
This is formalized by the following proposition, whose proof is Example~\ref{ex_agencycoord} in the Supplementary Material.
\begin{proposition}\label{prop_motivation}
	 The cumulative expected utility $U$ may be arbitrarily larger than the sum of the colluders' expected utilities when they participate in the ad auction individually.
\end{proposition}
%

%
%
In general, the media agency may adopt a \emph{randomized} bidding strategy in order to maximize $U$.
By letting $B^c$ be the set of all the possible colluders' bid profiles $b^c= (b^c_1, \ldots, b^c_{n_c})$, we denote by $\gamma \in \Gamma$ any randomized bidding strategy, where $\Gamma$ is the set of all the probability distributions over $B^c$.
%
%
Moreover, whenever $\gamma \in \Gamma$ has a finite support, we denote with $\gamma_{b^c}$ the probability of choosing a bidding strategy $b^c \in B^c$.

In this work, unless stated otherwise, we consider the case in which the bid profile $b^e= (b^e_1, \ldots, b^e_{n_e})$ of the external agents is drawn from a probability distribution $\gamma^e$.
%
%
Then, we define the expected revenue of bidder $i \in N_c$ for any bidding strategy $b^c \in B^c$ as $\tilde r_i(b^c) \coloneqq \mathbb{E}_{b^e \sim \gamma^e} r_i(b^c,b^e)$,
%
while their expected payment is as $\tilde \pi_i(b^c) \coloneqq \mathbb{E}_{b^e \sim \gamma^e} \pi_i(b^c,b^e)$.
%
%
%
In the rest of the paper, we assume that all algorithms have access to an oracle that returns the value of the expectations $\tilde r_i(b^c)$, and $\tilde \pi_i(b^c)$ for a bidding strategy $b^c\in B^c$ given as input.\footnote{Our results can be easily extended---only incurring in a small additive loss in cumulative expected utility---to the case in which the distribution $\gamma^e$ is unknown, but the algorithms have access to a black-box oracle that returns i.i.d.~samples drawn according to $\gamma^e$ (rather than returning expected values).}

%
The following Problem~\eqref{lp:problem1} encodes the maximization problem faced by the media agency, where the meaning of IC and IR constraints is described in the following.
%
\begin{subequations}\label{lp:problem1}
\begin{align}
 	\max_{q ,\gamma \in \Gamma} & \,\, \sum_{i \in N_c} \mathbb{E}_{b^c\sim \gamma}\mleft[\tilde r_i(b^c) - \tilde\pi_i(b^c)\mright] \quad \text{s.t.} \label{obj} \\
 	& \textnormal{IC}: \,\,\, \mathbb{E}_{b^c\sim \gamma}\mleft[\tilde r_i(b^c)\mright] - q_i \geq t_i  \hspace{1.3cm}  \forall i \in N_c  \label{c_incentive}\\
 	& \textnormal{IR}: \,\,\, \sum_{i \in N_c} q_i \geq \sum_{i \in N_c} \mathbb{E}_{b^c \sim \gamma}\mleft[\tilde \pi_i(b^c)\mright].    \label{c_transfer}
\end{align}
\end{subequations}

\noindent The elements of Problem~\eqref{lp:problem1} are defined as follows.

Objective~\eqref{obj} encodes the cumulative expected utility $U$ of the colluders and the media agency, in expectation with respect to the randomized bidding strategy $\gamma$.

Constraints~\eqref{c_incentive}---called \emph{incentive compatibility} (IC) constraints---ensure that the colluders are incentivized to be part of the media agency, rather than leaving it and participating in the ad auction as external agents (notice that these constraints naturally enforce bidders' individual rationality).
In particular, they guarantee that each colluder $i \in N_c$ achieves at least a minimum expected utility $t_i$, where the values $t_i \in [0,1]$ for $i \in N_c$ are given as input.\footnote{To the best of our knowledge, in the literature there is only one work by~\citeauthor{bachrach2010honor}~[\citeyear{bachrach2010honor}] that formalizes IC constraints for a setting that is similar to ours. \citeauthor{bachrach2010honor}~[\citeyear{bachrach2010honor}]~takes inspiration from the concept of core~\cite{cooperativegames} in cooperative games in order to define suitable IC constraints. However, this approach has many downsides. The most relevant issue of such an approach is that it is \emph{not} computationally viable, since computing the core would require exponential time in the number of colluders $n_c$.}

Constraint~\eqref{c_transfer} is an \emph{individual rationality} (IR) guarantee for the media agency.
Since the agency corresponds to the mechanism a payment $\sum_{i \in N_c} \mathbb{E}_{b^c \sim \gamma}[\tilde \pi_i(b^c)]$ in expectation over the clicks, the constraint requires that the sum of transfers $\sum_{i \in N_c} q_i$ covers the payment, so that the expected utility to the agency is positive.

In the following, we sometimes relax IC constraints by using $\delta$-IC constraints, for $\delta > 0$, which are defined as follows:
\begin{align}\label{c_eps_incentive}
& \delta\textnormal{-IC}: \,\,\, \mathbb{E}_{b^c\sim \gamma} \mleft[\tilde r_i(b^c)\mright] - q_i \geq t_i -\delta & \forall i \in N_c.
\end{align}

%
%
In the following, we call the scenario described so far, in which transfers $q_i$ could be negative, the \emph{arbitrary transfers} setting.
%
%
Monetary transfers from the media agency to the agents are \emph{not} always feasible in practice. 
Indeed, in some real-world scenarios a media agency could potentially loose customers by adopting a strategy for which some agents pay and some others are paid for participating in the same auction.
%
%
For these reasons, we introduce and study a second scenario, which we call \emph{transfers with limited liability} setting, where no agent is paid by the agency.
%
%
In such setting, Problem~\eqref{lp:problem1} is augmented with the following additional \emph{limited liability} (LL) constraints on the monetary transfers $q_i$:
\begin{align}
	& \textnormal{LL}: \,\,\, q_i \geq 0  & \forall i \in N_c.   \label{c_limited}
\end{align}
As we show in Section~\ref{sec:arbitrary}, there always exists an optimal solution to Problem~\eqref{lp:problem1} without LL constraints that is \emph{not} randomized.
The same does not hold for the problem with LL constraints, in which an optimal bidding strategy may be randomized, as in Example~\ref{ex_randombest} in the Supplementary Material.

We conclude by introducing the following assumption on the values $t_i$, guaranteeing that Problem~\eqref{lp:problem1} is feasible.\footnote{The proofs of all the results are in the Supplementary Material.}
%
%
\begin{assumption} \label{as_unica}
	There always exists a bidding strategy $b^c \in B^c$ such that $\tilde r_i(b^c) - \tilde \pi_i(b^c) \ge t_i $ for all $i \in N_c$.
	%
\end{assumption}
%
%
%
\begin{restatable}{proposition}{PropNonRand}
	When Assumption \ref{as_unica} holds, there always exists a non-randomized feasible solution to the Problem~\eqref{lp:problem1} with LL constraints.
\end{restatable}
In Appendix~\ref{appendixA} of the Supplementary Material, we provide some examples of natural choices for the values $t_i$, which arise from allocation and payment rules of the considered auction mechanisms and satisfy Assumption~\ref{as_unica}.
\section{Weighted Utility Problem}\label{sec:weighted}

In this section, we provide a polynomial-time algorithm for an abstract bid optimization problem 
we call \emph{weighted utility problem} (WUP).
%
This will be crucial in the following sections in order to solve Problem~\eqref{lp:problem1} with or without LL constraints.

Let $\mathcal{B}^c \coloneqq \{\bar{b}^c_1, \ldots, \bar{b}^c_d\}$ be a discrete set of $d$ different bid values, with $\bar{b}^c_1 \geq \ldots \geq \bar{b}^c_d$.
Moreover, given a bidding strategy $b^c \in B^c$ such that $b^c_i \in \mathcal{B}^c$ for all $ i \in N_c$, we write $b^c \in \bar{B}^c$ to express that each element of $b^c$ belongs to $\mathcal{B}^c$.

Then, WUP
reads as follows:
\begin{align} \label{obj_wup}
	\max_{b^c \in \bar{B}^{c}} \sum_{i \in N_c}\mleft( \hat{y}_i \, \tilde{r}_i(b^c) - \hat{x} \, \tilde{\pi}_i(b^c)\mright),
\end{align}
where $\hat{y}_i \ge 0$ for all $ i \in N_c$ and $\hat{x}\ge 0$.
A solution to Problem~\eqref{obj_wup} is a bidding strategy $b^c \in \bar{B}^c$ that maximizes the sum of suitable \emph{weighted} utilities of the colluders, which are defined so that colluder $i$'s expected revenue $\tilde{r}_i(b)$ is weighted by coefficient $\hat{y}_i$, while their expected payment $\tilde{\pi}_i(b)$ is weighted by coefficient $\hat{x}$.
In the following, for the ease of notation, for any $b^c \in \bar{B}^c$, we let $u_i^{\w}(b^c) \coloneqq \hat{y}_i \, \tilde{r}_i(b^c) - \hat{x} \, \tilde{\pi}_i(b^c)$ for every $i \in N_c$.
Notice that, when $\hat{y}_i=1$ for all $ i \in N_c$ and $\hat{x}=1$, then the objective of Problem~\eqref{obj_wup} coincides with the cumulative expected utility $U$.

\begin{figure}[!htp]
	\centering
	\includegraphics[scale=0.7]{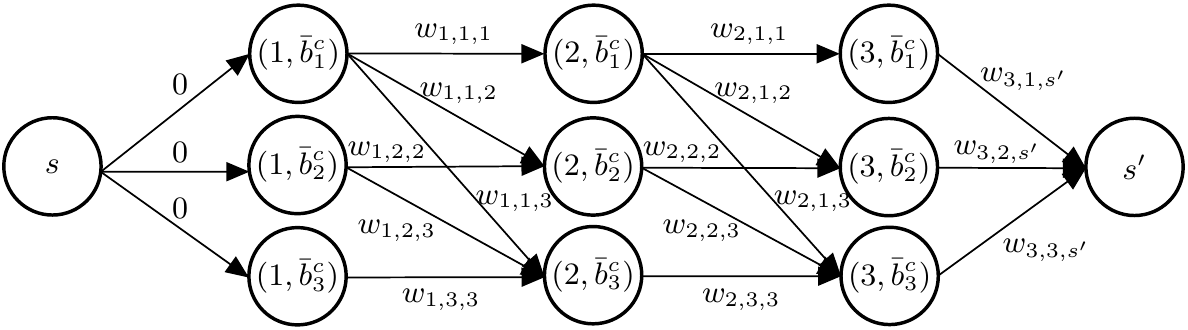}
	\caption{Structure of the graph when $N_c=\{1,2,3\}$ and $b^c_i \in \mathcal{B}^c=\{\bar{b}^c_1, \bar{b}^c_2, \bar{b}^c_3\}$ for all $ i \in N_c$.} 
	\label{fig:graph}
\end{figure}

As a first step, we consider Problem~\eqref{obj_wup} in which the profile of external agents' bids $b^e$ is fixed, which reads as follows: 
\begin{align} \label{obj_wup_fix}
	\max_{b^c \in \bar{B}^{c}} \sum_{i \in N_c} \mleft(\hat{y}_i \, r_i(b^c,b^e) - \hat{x} \, \pi_i(b^c, b^e)\mright).
\end{align}
%
For the sake of presentation, we first provide our results for Problem~\eqref{obj_wup_fix}, and, then, we show how they can be extended to Problem~\eqref{obj_wup}, where the bids of external agents are stochastic.

The main idea underpinning our results is to map Problem~\eqref{obj_wup} into a \emph{shortest path problem} \cite{10.1007/BF01386390}.  
Specifically, we show that the weighted utilities of bidding strategies are related to the length of paths in a particular directed acyclic weighted graph, whose structure and weights derive from the considered auction mechanism.
The following lemma is crucial for the construction of the graph.
%
\begin{restatable}{lemma}{OrderedBidding} \label{lem_bidorder}
	If the colluders in $N_c$ are ordered so that $\hat{y}_1 v_1 \geq \ldots \geq \hat{y}_{n_c} v_{n_c}$, then, for every $b^e$, any optimal bidding strategy $b^{c*} \in \bar{B}^c$ for Problem~\eqref{obj_wup_fix} is such that $b_1^{c*} \geq \ldots \geq b_{n_c}^{c*}$.
	%
\end{restatable}
%

%
%

We build the graph as follows (see Figure~\ref{fig:graph}).
\begin{itemize}
	\item The set of vertices $V$ contains $d n_c$ nodes, plus a source $s$ and a sink $s'$. There are $d$ nodes for each colluder, one for each possible bids. For $i \in N_c$ and $j \in [d]$, we let $(i,\bar{b}^c_j)$ be the node representing the bid $\bar{b}^c_j$ for bidder $i$.
	\item The set of arcs $A$ has cardinality $O(d^2n_c)$. The source $s$ is connected to nodes $(1,\bar{b}^c_j)$ and nodes $(n,\bar{b}^c_j)$ to the sink $s'$, for all $ j \in [d]$. For every $ i \in N_c \setminus \{1,n\}$ and $ j'\in [d]$, node $(i,\bar{b}^c_{j'})$ is connected to nodes $(i+1,\bar{b}^c_{j})$ with $j \in [d]: j \geq j'$. This implies that in each path the bids are ordered according to the colluders's indexes. This is w.l.o.g. by Lemma \ref{lem_bidorder}. In the following, we denote the arc from $(i,\bar{b}^c_{j'})$ to $(i+1,\bar{b}^c_{j})$ with the tuple $(i,\bar{b}^c_{j'},\bar{b}^c_{j})$.
	\item Each arc $(i,\bar{b}^c_{j'},\bar{b}^c_{j})$ has a weight $w_{i,j',j}$.
	For $i \in N_c \setminus \{n\}$ and $j',j \in [d]$, $w_{i,j',j}$ is a fraction of the cumulative utility, once fixed the bids $b_{i'}^c$ with $i'<i$, $b^c_i = \bar{b}^c_{j'}$, and $b^c_{i+1}=\bar{b}^c_j$.
	We analogously denote the weights on arcs from nodes $(n,\bar{b}^c_{j'})$ to the sink $s'$ by $w_{n,j',s'}$, while those on arcs from the source $s$ to agent $1$'s nodes are $w_{s,j}=0$, for all $ j \in [d]$. 
\end{itemize}
%

%
A \emph{directed path} $\sigma \in \Sigma$ is a sequence of arcs connecting the source $s$ to the sink $s'$. The \emph{length} $W_\sigma$ of path $\sigma$ is the sum of the weights of the arcs in the path:
\begin{align}
	W_\sigma = \sum_{(i,\bar{b}^c_{j'},\bar{b}^c_{j}) \in \sigma} w_{i,j',j}.
\end{align}
The shortest path problem on the weighted graph is
\begin{align} 
	\min_{\sigma \in \Sigma} - W_\sigma \label{obj_short}.
\end{align}
We let $n(b^c_i) \coloneqq \sum_{j=1}^{n_e} \mathbbm{1}_{\{b^e_j > b^c_i\}}$ be the number of external agents with a bid larger than agent $i$'s bid.
%

In the GSP mechanism, we define the weight of $(i,\bar{b}^c_{j'},\bar{b}^c_{j})$ as 
\begin{align}
w_{i,j',j} = &\lambda_{i+n(b^c_i)} \Big(\hat{y}_i v_i  - \hat{x}\max\Big\{ b^c_{i+1},\max_{b^e_j < b^c_i}{b^e_j}\Big\}\Big),
\end{align}
where $b^c_{i} = \bar{b}^c_{j'}$ and $b^c_{i+1} = \bar{b}^c_{j}$.
%
%
Notice that, under GSP payments, the weight $w_{i,j',j}$ is equal to $u_i^{\w}(b^c)$ when colluder $i$ is assigned to the $(i+n(b^c_i))$-th slot.
%
%

In the VCG mechanism, we define the weights as follows 
\begin{align}
	&w_{i,j',j} = \hat{y}_i r_{i}(b^c, b^e) - \hat{x} \Big[ g_i(b^c, b^e) + \ell_i(b^c, b^e) \Big], 
\end{align}
where
\begin{align}
	&r_{i}(b) = \lambda_{i+n(b^c_i)} v_i, \\
	& g_i(b) =  (i-1) b^c_i (\lambda_{i + n(b^c_i) - 1} - \lambda_{i + n(b^c_i)}),\\
	&\ell_i(b)=  \sum_{b^e_h \in (b^c_{i+1},b^c_i]} i \, b^e_h (\lambda_{h+i-1} - \lambda_{h + i}),
\end{align}
$b^c_{i} = \bar{b}^c_{j'}$, and $b^c_{i+1} = \bar{b}^c_{j}$. 
%
In the VCG mechanism, $w_{i,j',j}$ has a less intuitive interpretation than for the GSP mechanism.
In particular, $w_{i,j',j}$ is composed of a revenue term, which is agent $i$'s expected revenue $r_i(b)=\lambda_{i+n(b^c_{i})} v_i$, weighted by $\hat{y}_i$, and two payment terms, $g_i(b)$ and $\ell_i(b)$, weighted by $\hat{x}$.
The latter are two parts of the cumulative payment from the agency to the mechanism, related to the externalities of agents $i' \in N_c$, with $i' <i$, when agents $i$ and $i+1$ bid $b^e_{j'}$ and $b^e_{j}$, respectively.
In particular, the fraction of the expected payment $\pi_{i'}(b)$ related to agent $i'$, with $i'<i$, due to the presence of agent $i$ bidding $b^c_i$, is $b^c_i (\lambda_{i + n(b^c_i) - 1} - \lambda_{i + n(b^c_i)})$. Term $g_i(b)$ is the summation of such payments over all agents $i'$. Moreover, agent $i'$'s fraction of expected payment due to the presence of external agents bidding $b^e_h$ is $b^e_h (\lambda_{h+i-1} - \lambda_{h + i})$. Term $\ell_i(b)$ is the summation of such expected payments due to external agents with bid $b^e_h\in (b^c_{i+1},b^c_i]$.

The following lemma establishes the relation between the length of the paths in the graph and the objective of Problem~\eqref{obj_wup_fix}.
\begin{restatable}{lemma}{MapPath}\label{lem_map}
	Consider a path $\sigma \in \Sigma$ composed by the sequence of nodes $\{(1,\bar{b}^c_j), \ldots, (n,\bar{b}^c_{j'})\}$. Then,
	\begin{align} \label{eq_pathbid}
		W_\sigma = \sum_{i \in N_c} \hat{y}_i r_i(b^c, b^e) - \hat{x} \pi_i(b^c, b^e),
	\end{align}
which is the weighted utility in Problem~\eqref{obj_wup_fix} with fixed external bids $b^e$ and bidding strategy $b^c = (\bar{b}^c_j, \ldots, \bar{b}^c_{j'}) \in \bar{B}^c$. Moreover, for each bidding strategy $b^c = (\bar{b}^c_j, \ldots, \bar{b}^c_{j'}) \in \bar{B}^c$ there exists a corresponding path $\sigma \in \Sigma$ composed by the sequence of nodes $\{(1,\bar{b}^c_j), \ldots, (n,\bar{b}^c_{j'})\}$.
\end{restatable}
%
%

The following theorem provides the polynomial-time algorithm solving Problem~\eqref{obj_wup_fix}, which works by simply finding a shortest path of the graph.
\begin{restatable}{theorem}{polyweight}\label{th_poly_1}
	There exists a polynomial-time algorithm which solves Problem~\eqref{obj_wup_fix}. 
\end{restatable}


Finally, by substituting the quantities involved in Problem~\eqref{obj_wup_fix} with their expectations, thanks to the linearity of the objective we get that:
\begin{restatable}{theorem}{PolyExpectation}\label{th_poly}
	There exists a polynomial-time algorithm which solves Problem~\eqref{obj_wup}.
\end{restatable}

\section{Arbitrary Transfers Setting}\label{sec:arbitrary}

In this section, we provide an approximate solution to the media agency problem with arbitrary transfers.
In particular, we design a bi-criteria FPTAS that returns solutions providing an arbitrary small loss $\varepsilon > 0$ with respect to the optimal value of the problem, by relaxing the IC constraints by the additive factor $\varepsilon$.
As a first step, we show that there always exists a non-randomized solution to Problem~\eqref{lp:problem1}.

\begin{restatable}{lemma}{DetSolution}\label{lem_deterministic}
	In the arbitrary transfers setting, there always exists an optimal non-randomized solution to Problem~\eqref{lp:problem1}.
\end{restatable}

Then, we show how to reduce the problem to a new one working with a finite (discretized) set of bids, in order to apply the results provided in Section~\ref{sec:weighted}.
As a first result, given a probability value $p\in [0,1]$ and a minimum discretization step $\eta \in [0,1]$, we show how to split the space of bids $[0,1]$ into a suitably-defined set of intervals using the recursive algorithm whose pseudo-code is provided in Algorithm~\ref{alg}.\footnote{In Algorithm~\ref{alg}, we assume that we can compute some probabilities exactly. It is easy to extend our results---with only a small approximation error---to the case in which the algorithm has access only to samples of the probability distribution over the external bids.}

\begin{algorithm}[!htp] 
	\caption{$\textsc{Rec}((\alpha,\beta],p,\eta)$}\label{alg}
	\textbf{Input:} interval $(\alpha,\beta]\subseteq [0,1]$; probability $p\in [0,1]$; minimum discretization step $\eta \in [0,1]$
	\begin{algorithmic}[1]
		 \IF{$\underset{b^e \sim \gamma^e}{\mathbb{P}} \left\{ \exists j \in N_e: b^e_j \in (\alpha,\beta] \right\}\le p \vee \beta-\alpha \le \eta$} \label{line}
		\RETURN $\{(\alpha,\beta]\}$
	   \ELSE
	   \STATE { $\mathcal{I}_\text{L} \gets \textsc{Rec}\left( \left(\alpha,\frac{\alpha+\beta}{2}\right],p,\eta \right)$}
	   \STATE {$\mathcal{I}_\text{R} \gets \textsc{Rec}\left( \left( \frac{\alpha+\beta}{2}, \beta \right],p,\eta \right)$}
	   \RETURN $\mathcal{I}_{L} \cup \mathcal{I}_\text{R}$
	   \ENDIF
	\end{algorithmic} 
\end{algorithm}

We prove the following:
\begin{restatable}{lemma}{REC}\label{lem_REC}
	 Given $p \in [0,1]$ and $\eta \in [0,1]$, $\textsc{Rec}((0,1], p,\eta)$ returns a set $\{(\alpha_j,\beta_j]\}_{j \in [k^*]}$ composed of $k^*\le \frac{2 n_e}{p} \log{ \frac{1}{\eta}}$ intervals such that, for every interval $(\alpha_j,\beta_j]$, it holds either $\mathbb{P}_{b^e \sim \gamma^e} \left\{ \exists i \in N_e: b^e_i \in (\alpha_j,\beta_j] \right\} \le p$ or $\beta_j-\alpha_j\le \eta$.
	 Moreover, it holds that $\bigcup_{j \in [k^*]} (\alpha_j,\beta_j] =(0,1]$ and the procedure runs in time polynomial in $n_e$, $\frac{1}{p}$, and $\log \frac{1}{\eta}$.
\end{restatable}

Let $I^{p,\eta} \coloneqq \{(\alpha_j,\beta_j]\}_{j \in [k^*]}$ be the set of intervals returned by $\textsc{Rec}((0,1], p,\eta)$.
The next step is to show that, for $\eta$ small enough, $\mathbb{P}_{b^e \sim \gamma^e} \left\{ \exists i \in N_e: b^e_i \in (\alpha_j,\beta_j) \right\}\le p$ for every interval $(\alpha_j,\beta_j]$.
This holds by definition for each interval $(\alpha_j,\beta_j]$ with $\beta_j-\alpha_j>\eta$.
Thus, let us consider the intervals $(\alpha_j,\beta_j]$ such that $\beta_j-\alpha_j\le \eta$.
Let $M$ be the maximum number of bits needed to represent the bids $b^e_i$ that are in the support of probability distribution $\gamma^e$.
By setting $\eta=2^{-M}$, we have that all the bids $b^e_i$ in the interval $(\alpha_j,\beta_j]$ are equal to $\beta_j$.\footnote{Our algorithm runs in time logarithmic in $\frac{1}{\eta}$ and hence polynomial in the size of the binary representation of bids $b^e$.}
Hence, it holds $\mathbb{P}_{b^e \sim \gamma^e} \left\{ \exists i \in N_e: b^e_i \in (\alpha_j,\beta_j) \right\}=0$.

Now, we define the set of discretized bidding strategies $B^{c,p}$ employed in the rest of the paper, as $B^{c,p} \coloneqq \bigcup_{(\alpha,\beta] \in I^{p,\eta}} \bigcup_{i \in N_c}\{\alpha+\tau i\}$ for a small $\tau > 0$ and $\eta=2^{-M}$.\footnote{In the following, we ignore the loss in cumulative expected utility that results from the introduction of $\tau > 0$. Notice that this parameter is only necessary to induce specific tie-breaking rules and our results can be easily extended to consider the loss in utility due to $\tau$. Moreover, $\tau$ can be taken to be exponentially small in the size of the problem instance, and hence negligible.}
Next, we show that we can restrict to bid profiles $B^{c,p}$ with a small loss in utility and by relaxing IC constraints.

First, we provide the following auxiliary result.
\begin{restatable}{lemma}{DiscretizedBid}\label{lm:discretized}
	Given $p \in [0,1]$, for any bidding strategy $b^c \in B^c$, there exists a discretized bidding strategy $\hat b^c \in B^{c,p}$:
	\begin{itemize}
		\item $ \tilde \pi_i(\hat b^c) \le \tilde \pi_i( b^c)  $ for all $ i \in N_c$; and
		\item $ \tilde r_i(\hat b^c) \ge \tilde r_i( b^c) -p $ for all $  i \in N_c $.
	\end{itemize}
\end{restatable}

Then, by exploiting Lemma~\ref{lm:discretized} we can prove the following Lemma~\ref{lm:arbitrary}.
Intuitively, the lemma shows that, given a probability $p\in [0,1]$ and an optimal discretized  bidding strategy $\hat{b}^c \in \argmax_{b^c \in B^{c,p}} \sum_{i \in N_c} \tilde r_i(b^c) - \tilde \pi_i(b^c)$, one can find an approximate solution to Problem~\ref{lp:problem1} in polynomial time.
%
%
\begin{restatable}{lemma}{Arbitrary} \label{lm:arbitrary}
	Given $p\in [0,1]$ and an optimal discretized bidding strategy $\hat b^{c} \in \argmax_{b^c \in B^{c,p}}  \sum_{i \in N_c} \tilde r_i(b^c) - \tilde \pi_i(b^c)$, we can find in polynomial time a $p$-IC and IR solution to Problem~\eqref{lp:problem1} with value at least $OPT-p n_c$, where $OPT$ is the optimal value of Problem~\eqref{lp:problem1}.
\end{restatable}

By Theorem~\ref{th_poly}, for any probability $p \in [0,1]$, it is possible to find an optimal discretized bidding strategy $\hat b^{c} \in \argmax_{b^c \in B^{c,p}  }  \sum_{i \in N_c} \tilde r_i(b^c) - \tilde \pi_i(b^c)$ in time polynomial in the instance size and in $\frac{1}{p}$, since, as it is easy to check, the number of possible discretized bids in $B^{c,p}$ is polynomial in $\frac{1}{p}$.
Moreover, by employing Lemma~\ref{lm:arbitrary}, we can use the bidding strategy $\hat b^{c}$ to find an approximated solution to Problem~\eqref{lp:problem1} in polynomial time.
Hence, given any $\varepsilon>0$, it is sufficient to choose $p \in [0,1]$ so that $\frac{1}{p} \in \text{poly}(\frac{1}{\varepsilon},n)$ in order to obtain an $\varepsilon$-IC and IR approximate solution to Problem~\ref{lp:problem1}, as stated by the following theorem.
\begin{theorem}
	Given $\varepsilon>0$, there exists an algorithm that runs in time polynomial in the instance size and $\frac{1}{\varepsilon}$, which returns an $\varepsilon$-IC and IR solution to Problem~\eqref{lp:problem1} with value at least $OPT-\varepsilon$, where $OPT$ is the optimal value of Problem~\eqref{lp:problem1}.
\end{theorem}

\section{Transfers with Limited Liability Setting}

In this section, we provide an approximate solution to the media agency problem with limited liability constraints.
In particular, similarly to the arbitrary transfers setting, we design a bi-criteria FPTAS that returns solutions providing an arbitrary small loss $\varepsilon > 0$ with respect to the optimal value of the problem, by relaxing the IC constraints by $\varepsilon$.

As a first step, we show that we can restrict Problem~\eqref{lp:problem1} with LL constraints to work with the set $B^{c,p}$ by only incurring in a small loss in the objective function value and IC constraints satisfaction.
Notice that, Problem~\eqref{lp:problem1} with LL constraints restricted to discretized bids $B^{c,p}$ does not only have a smaller optimal value than Problem~\eqref{lp:problem1} with LL constraints, but it can also result in an infeasible problem, since Assumption~\ref{as_unica} is \emph{not} necessarily satisfied for a discretized bidding strategy $b^c \in B^{c,p}$.
However, we can prove that, given a probability $p \in [0,1]$, the following LP~\eqref{lp:Limit} that uses only bids in $B^{c,p}$ and relaxes the IC constraints of quantity $p$ is feasible and has value at least $OPT-pn_c$, where $OPT$ is the optimal value of Problem~\eqref{lp:problem1} with LL constraints.
\begin{subequations}\label{lp:Limit}
\begin{align} 
\max_{q \geq 0, \gamma \in \Delta_{B^{c,p}}} &\sum_{b^c \in B^{c,p}} \gamma_{b^c} \sum_{i \in N_c}  \tilde{r}_i(b^c) - \tilde{\pi}_i(b^c)  \label{obj_lim2} \quad  \textnormal{s.t.}\\
 &\hspace{-1cm}\sum_{b^c \in B^{c,p}} \gamma_{b^c} \tilde r_i(b^c) - q_i \geq t_i -p  \qquad \forall i \in N_c \label{c_inc2}\\
 &\hspace{-1cm}\sum_{i \in N_c} q_i \geq \sum_{b^c \in B^{c,p}} \gamma_{b^c} \sum_{i \in N_c} \tilde \pi_i(b^c). \label{c_transf2}
\end{align}
\end{subequations}

Formally, we prove the following:
\begin{restatable}{lemma}{Feasible} \label{lm:feasible}
	LP~\eqref{lp:Limit} is feasible.
	Moreover, the optimal value of LP~\eqref{lp:Limit} is at least $OPT-pn_c$, where $OPT$ is the optimal value of Problem~\eqref{lp:problem1} with LL constraints.
\end{restatable}

In the rest of the section, we will provide an algorithm to solve LP~\eqref{lp:Limit} by using the ellipsoid method.
In order to do that, we use the dual LP~\eqref{lp:dual}, in which variables $y=\{y_1, \ldots,y_{n_c}\}$, $x$, and $z$ are related to Constraints~\eqref{c_inc2},~\eqref{c_transf2}, and the constraint $\gamma \in \Delta_{B^{c,p}}$, respectively.
\begin{subequations} \label{lp:dual}
\begin{align} 
	\min_{y \leq 0,x,z} & \,\, \sum_{i \in N_c} (t_i-p) y_i + z  \quad \textnormal{s.t.} \label{obj_lim_dual} \\
	 &\sum_{i \in N_c} y_i \tilde r_i(b^c) - x \sum_{i \in N_c} \tilde \pi_i(b^c) + z \geq \nonumber \\
	& \qquad  \sum_{i \in N_c} \tilde r_i(b^c)-\tilde \pi_i(b^c) \quad\quad\quad   \forall b^c \in B^{c,p} \label{dual:2}\\
	&-y_i +x \geq 0   \hspace{3cm}\forall i \in N_c. \label{dual:3}
\end{align}
\end{subequations}

By Lemma~\ref{lm:feasible}, the primal LP~\eqref{lp:Limit} is feasible (and bounded), and, thus, it holds strong duality.
As a consequence, in order to provide a polynomial-time algorithm to solve LP~\eqref{lp:Limit}, it is enough to apply the ellipsoid method to the dual LP~\eqref{lp:dual}, which can be done in polynomial time since the latter has polynomial-many variables and exponentially-many constraints.
This is possible by providing a polynomial-time separation oracle that, given an assignment of values to the variables as input, returns a violated constraint (if any).
Since there are only polynomially-many Constraints~\eqref{dual:3}, we can check if one of them is violated in polynomial time.
Moreover, in order to find whether there exists a violated Constraint~\eqref{dual:2}, it is sufficient to solve the weighted utility problem in Equation~\eqref{obj_wup} by setting $\hat{y_i} =(1-y_i)$ for each $i \in N_c$ and $\hat x=x-1$.
By Theorem~\ref{th_poly}, this can be done in polynomial time with computing a shortest path of a suitable graph.
Hence, we can prove the following theorem.
\begin{restatable}{theorem}{ExistPloyAlgorithm}
		Given $\epsilon>0$, there exists an algorithm that runs in time polynomial in the instance size and $\frac{1}{\varepsilon}$ and returns an $\varepsilon$-IC and IR solution to Problem~\eqref{lp:problem1} with LL constraints having value at least $OPT-\varepsilon$, where $OPT$ is the optimal value of Problem~\eqref{lp:problem1} with LL constraints.
\end{restatable}

\clearpage
\bibliographystyle{named}
\bibliography{ijcai22}

\clearpage
\onecolumn
\begin{center}
	\LARGE{\textbf{Supplementary Material}}
\end{center}
\appendix

\section{Appendix -- Examples}
\label{appendixA}

\begin{example} \label{ex_agencycoord}
	There are two colluders $N_c = \{1,2\}$, with valuations $v_1 = v_2 + \varepsilon$ and $v_2 \in [0,1]$ for $\varepsilon >0$.
	There is one slot $M=\{1\}$, with click through rate $\lambda_1=1$, and one external agent whose bid is $b_1^e=0$.
	Notice that, in one-slot settings, VCG and GSP mechanisms define the same payments and, thus, they are both truthful.
	We consider the following cases:
	\begin{itemize}
		\item \textnormal{Without media agency:} the colluders' bid profile is $b^c=(b^c_1,b^c_2) = (v_1,v_2)$ and agent $1$ wins the slot paying $v_2$ (since $b^c_1 > b^c_2 > b^e_1$); then, the cumulative expected utility of the colluders is $U_{\textnormal{w/o}} = u_1 + u_2 = \varepsilon$.
		\item \textnormal{With media agency:} the colluders' bid profile is $b^c=(v_1,0)$ and agent $1$ wins the slot paying $0$; then, the cumulative expected utility is $U_{\textnormal{w}}=u_{1} + u_2 = v_1 = v_2 + \varepsilon$.
	\end{itemize}
	Thus, by letting $\varepsilon \rightarrow 0$, we have that $\frac{U_{\textnormal{w}}}{U_{\textnormal{w/o}}} \rightarrow +\infty$.
	%
\end{example}

\begin{example}\label{ex_agencycoord_at}
	In Example \ref{ex_agencycoord} we consider the deterministic bidding strategies $b^c=(v_1,0)$ played with probability $\gamma_{b^c}=1$, and $b^e=0$ with $\gamma_{b^e}=1$. Agent $2$ has the same utility $u_2=0$ with or without the coordination of the agency. Moreover, agent $2$ is incentivized to stay under the agency when $t_2=0$. However, suppose that $t_2=\delta$ and $t_1=\varepsilon$ (\emph{i.e.}, agent $1$'s utility without the agency). In this case, the optimal expected utility $U=v_1$, achieved by the strategy $b^c=(v_1,0)$, should be partitioned between the agents s.t. $u_2(b) \geq \delta$. A feasible solution is $q_1=\delta$ and $q_2=-\delta$ (recall that $\pi_1(b)=\pi_2(b)=0$). The corresponding utilities are $u_1(b) = r_1(b)-q_1=v_1-\delta$ for agent $1$ and $u_2(b) = r_2(b)-q_2=\delta$ for agent $2$. Under the hypothesis that $v_1-\delta \geq \varepsilon$, this is a feasible solution maximizing the objective function (\ref{obj}).
\end{example}

\begin{example}\label{ex_randombest}
	In this example the best non-randomized bidding strategy provides a lower utility w.r.t. the best randomized bidding strategy.
	Consider a set of two slots $M=\{1,2\}$ with click through probabilities $\lambda_1 \geq \lambda_2$, a set of two colluders $N_c=\{1,2\}$ and a set of one external agent $N_e=\{1\}$. It holds that $b^e_1 \leq v_2 \leq v_1$. The ties are broken in favor of the colluders. Suppose that $b^c_1=b^c_2=1$, $b^e= \frac{2}{3}$, $\lambda_1=\lambda_2=1$, $t_1 = \delta_1$, and $t_2=\delta_2$, with $\delta_1$ and $\delta_2$ arbitrarily small.
	Consider the non-randomized bidding strategy $b^c_1=v_1$, $b^c_2=b^e$.
	Consider the randomized bidding strategy: play with probability $\gamma_{b^{c'}} = \frac{1}{2}$ stategy $b^{c'}_1=\epsilon$, $b^{c'}_2=0$, with small $\epsilon$ and play with probability $\gamma_{b^{c''}} = \frac{1}{2}$ stategy $b^{c''}_1=0$, $b^{c''}_2=\epsilon$, with small $\epsilon$.
	Under both VCG and GSP mechanisms we have that:
	\begin{itemize}
		\item The non-randomized strategy provides cumulative utility $U^{NR} = u_1(b^c) + u_1(b^c) = \frac{1}{3} + \frac{1}{3} = \frac{2}{3}$.
		\item The randomized strategy: bidding strategy $b^{c'}$ provides cumulative utility $U^{R'}=1$, and $b^{c'}$ provides $U^{R''}=1$. The expected cumulative utility on the stochasticity of $b^c$ is $U^R=1$.
		\item $U^R > U^{NR}$
	\end{itemize}
\end{example}


\section{Appendix -- How to define the value of leaving the agency}\label{ap_setting_t}
Consider a VCG mechanism and fix $t_i$ as the utility of agent $i$, $\forall{i \in N^c}$, when all agents in $N$ paritcipate to the auction without the coordination of the agency. With such choice of parameter $t_i$, $\forall{i \in N^c}$, a feasible solution to the optimization problem \eqref{lp:problem1} with LL always exists and is given by the bidding strategy $b^c=(v_1, \ldots, v_{n^c})$.  

For the case of GSP mechanism, there are many possible choices of value $t_i$ as there is no dominant strategy. For instace, consider the Balanced Bidding strategy from \citeauthor{cary2007greedy}~[\citeyear{cary2007greedy}]. Parameter $t_i$ can be set as the expected utility of agent $i$, when the colluders adopt such strategy considering the external bids in expectetion. 
\section{Appendix -- Proofs}

\PropNonRand*
\begin{proof}
	Suppose that there exists a bidding strategy $b^c \in B^c$ such that $\tilde r_i(b^c) - \tilde \pi_i(b^c) \ge t_i $ for all $i \in N_c$. We prove the proposition by showing that Problem~\eqref{lp:problem1} has a feasible solution composed by the non-randomized bidding strategy $b_c$ and transfers $q_i = \tilde \pi_i(b^c)$ for all $i \in N_c$. 
	Constraints \eqref{c_incentive} are satisfied because of the condition $\tilde r_i(b^c) - \tilde \pi_i(b^c) \ge t_i $ for all $i \in N_c$.
	Then, by substituting $q_i = \tilde \pi_i(b^c)$ in Constraint \eqref{c_transfer}, we show that also the IR constraint is satisfied.
	We conclude the proof observing that under GSP and VCG mechanisms $\tilde \pi_i(b^c)>0$ for all $i \in N_c$, therefore also LL constraints are satisfied.
\end{proof}

\OrderedBidding*
\begin{proof}
	We prove the result in $\GSP$ auctions.
	Consider two colluders, \emph{i.e.} $N_c=\{1, \ldots, n_c\}$, having weighted valuations ${v}^{\w,c}_i=\hat{y}_i v_i^c$ and $v^{\w,c}_j=\hat{y}_jv_j^c$ such that $v^{\w,c}_i \geq v^{\w,c}_j$.
	Consider an external bid vector $b^e$ and a set of slot $M=\{1,\ldots,m\}$ such that their click through rates are $\lambda_1 > \ldots> \lambda_m$.
	Suppose that ties are broken in favor of the colluders and fix a bidding strategy $b^c=(b^c_1, \ldots, b^c_{n_c})$, such that $b^c_1 \geq \ldots \geq b^c_h \geq \ldots \geq b^c_k \geq \ldots \geq b^c_{n_c}$.
	Consider $\GSP$ allocation function $f$ such that $f(i)=j$ if agent $i$ is allocated to slot $j$.  
	The expected cumulative utility is
	\begin{align}
		U = \sum_{i \in N_c} U_i &= U_h + U_k + \sum_{i \in N_c \setminus \{h,k\}} U_i \nonumber \\
		& = \lambda_{f(h)} (v^{\w,c}_h - \max\{b^c_{h+1}, \max_{b^e_j \leq b^c_h} b^e_j\}) +  \lambda_{f(k)} (v^{\w,c}_k - \max\{b^c_{k+1}, \max_{b^e_j \leq b^c_k} b^e_j\})  \nonumber \\
		& + \sum_{i\in N_c \setminus \{h,k\}}  \lambda_{f(i)} (v^{\w,c}_i - \max\{b^c_{i+1}, \max_{b^e_j \leq b^c_i} b^e_j\}) \nonumber  \\
		& \geq \lambda_{f(h)} (v^{\w,c}_k - \max\{b^c_{h+1}, \max_{b^e_j \leq b^c_h} b^e_j\}) +  \lambda_{f(k)} (v^{\w,c}_h - \max\{b^c_{k+1}, \max_{b^e_j \leq b^c_k} b^e_j\})  \nonumber \\
		& + \sum_{i\in N_c \setminus \{h,k\}}  \lambda_{f(i)} (v^{\w,c}_i - \max\{b^c_{i+1}, \max_{b^e_j \leq b^c_i} b^e_j\}) \nonumber  \\
		& = \hat{U}_h + \hat{U}_k + \sum_{i \in N_c \setminus \{h,k\}} \hat{U}_i = \hat{U}, \nonumber 
	\end{align}
	where $\hat{U}$ is the utility provided by any bidding strategy in which bid $b^c_h$ is switched with bid $b^c_k$.	This proves that the maximum utility is reached through a bidding strategy ordered in non-increasing values of $b^c_i$.
	
	The result can be proved for $\VCG$ auction mechanism by analogous steps.
\end{proof}

\MapPath*
\begin{proof}
	We prove that every path in the graph corresponds to a specific bidding strategy in the auction such that the sum of the weights $W_\sigma$ of the arcs in the path is the weighted utility in the optimization Problem~\eqref{obj_wup_fix} fixed that strategy, and viceversa. 
	We show the existence of a bidding strategy given a path, of a path given a bidding strategy, and how to find them. The procedure follows from the construction of the graph.
	\begin{itemize}
		\item Given a path $\{(1,\bar{b}^c_j), \ldots, (n_c,\bar{b}^c_{j'})\}$, the corresponding bidding strategy is $b^c_1=\bar{b}^c_j, \dots, b^c_{n_c}=\bar{b}^c_{j'}$.
		\item Given a bidding strategy $b^c_1=\bar{b}^c_j, \dots, b^c_{n_c}=\bar{b}^c_{j'}$, such that w.l.g. $b^c_1 \geq \ldots \geq b^c_{n_c}$ by Lemma~\ref{lem_bidorder},  the corresponding path is $\{(1,\bar{b}^c_j), \ldots, (n_c,\bar{b}^c_{j'})\}$, which always exists by the way in which the graph is built.
	\end{itemize}
	Now, we prove Equation (\ref{eq_pathbid}) under GSP and VCG mechanisms.
	
	In GSP auction, the weight $w_{i,j',j}$ is $u_i^{\w}(b^c)$ under GSP payments when colluder $i$ is assigned to the $(i+n(b^c_i))$-th slot:
	\begin{align}
		W_\sigma = \sum_{(i,\bar{b}^c_{j'},\bar{b}^c_j) \in \sigma} w_{i,j',j} = \sum_{i \in N_c} \hat{y}_i r_i(b^c) - \hat{x} \pi_i^{\GSP}(b^c) \label{gsp_weights}
	\end{align}
	The equivalence between the solutions of the shortest path (\ref{obj_short}) and Problem~\eqref{obj_wup_fix} follows from equality (\ref{gsp_weights}) and from the fact that problem (\ref{obj_short}) is equivalent to 
	\begin{align}
		\max_{\sigma \in \Sigma} W_\sigma .
	\end{align}
	
	
	Now, we analyze the case of VCG mechanism.
	We define $n(b^e_j)= \sum_{k=1}^{n_c} \mathbbm{1}_{\{b^c_k \geq b^e_j\}}$ the number of colluders with a bid larger than or equal to external bid $b^e_j$.
	We have:
	\begin{align}
		W_\sigma = &\sum_{(i,\bar{b}^c_{j'},\bar{b}^c_j) \in \sigma} w_{i,j',j} \\
		=& \sum_{i \in N_c} \mleft( \hat{y}_i r_i(b_c) - \hat{x}\mleft( (i-1) b^c_i (\lambda_{i + n(b^c_i) - 1} - \lambda_{i + n(b^c_i)}) - \sum_{b^e_h \in (b^c_{i+1},b^c_i]} i \, b^e_h (\lambda_{h+i-1} - \lambda_{h + i}) \mright) \mright) \\
		= & \sum_{i \in N_c} \mleft(\hat{y}_i r_i(b_c) - \hat{x} \mleft( \sum_{i' >i} b^c_{i'}(\lambda_{i' + n(b^c_{i'}) - 1} - \lambda_{i' + n(b^c_{i'})}) -\sum_{b^e_j \leq b^c_i} b^e_j (\lambda_{j+n(b^e_j)-1} - \lambda_{j + n(b^e_j)})\mright)\mright) \label{eq_vcg_bcbe}\\
		= & \sum_{i \in N_c} \mleft(\hat{y}_i r_i(b_c) - \hat{x}\sum_{j=i+1}^{m+1} b_{j}(\lambda_{j-1} -\lambda_{j} )\mright) \label{eq_vcg_b}\\
		= & \sum_{i \in N_c} \mleft(\hat{y}_i r_i(b_c) - \hat{x}\pi_i^{\VCG}(b^c)\mright)
	\end{align}
	where, for the ease of notation, we let $\lambda_{m+1} = 0$. 
	Notice the change of notation from Equation (\ref{eq_vcg_bcbe}) to Equation (\ref{eq_vcg_b}).
	In Equation (\ref{eq_vcg_bcbe}) we use bidding strategy $b^c = (b^c_1, \ldots , b^c_{n_c})$ and external agents' bids profile $b^e = (b^e_1, \ldots , b^e_{n_e})$, while in Equation (\ref{eq_vcg_b}) we use the bid profile of all agents $b = (b^c,b^e)= (b_1, \ldots , b_{n})$, where bids are ordered by decreasing value of $b_i$.
\end{proof}

\PolyExpectation*
\begin{proof}
	It suffices to solve the shortest path problem by assigning to each arc $(i,\bar{b}^c_{j'}, \bar{b}^c_{j})$ the weight computed in expectetion w.r.t. the distribution of $b^e$, which we denote by $\tilde{w}_{i,j',j} = E_{b^e \sim \gamma^e_{b^e}}[w_{i,j',j}]$. 
	The result follows from the linearity of the objective function of the Problem~\eqref{obj_wup_fix}.
	\begin{align}
		&\sum_{(i, \bar{b}^c_{j'},\bar{b}^c_j) \in p} \tilde{w}_{i,j',j} = \\
		&\sum_{(i, \bar{b}^c_{j'},\bar{b}^c_j) \in p} E_{b^e \sim \gamma^e_{b^e}}[w_{i,j',j}] = \\
		&E_{b^e \sim \gamma^e_{b^e}}\mleft[ \sum_{(i, \bar{b}^c_{j'},\bar{b}^c_j) \in p} w_{i,j',j}\mright] = \\
		&E_{b^e \sim \gamma^e_{b^e}}\mleft[ \sum_{i \in N_c}    \hat{y}_i r_i(b^c) - \hat{x} \pi_i(b^c)    \mright] = \\
		&\sum_{i \in N_c} \hat{y}_i \tilde{r}_i(b^c) - \hat{x} \tilde{\pi}_i(b^c) 
	\end{align}
\end{proof}

\DetSolution*
\begin{proof}
	We provide an optimal solution $\gamma^*, q$ to LP \eqref{lp:problem1} such that $\gamma^*_{b^{c,*}}=1$ for a bid profile $b^{c,*}$.
	Let $b^{c,*} \in \argmax_{b^c \in B^c} \sum_{i \in N_c} \tilde{r}_i(b^c) - \tilde{\pi}_i(b^c)$.
	Moreover, let $q_i=  \tilde r_i(b^{c,*})-t_i$.
	By the optimality of $\hat b^c$, the value of the objective \eqref{obj} is optimal.
	Moreover, Constraints \eqref{c_incentive} are satisfied by construction.
	To conclude the proof, we show that constraint \eqref{c_transfer} is satisfied.
	In particular $\sum_{i \in N_c} [q_i-\tilde \pi_i(b^{c,*})]=\sum_{i \in N_c} ( \tilde r_i(b^{c,*}) - \tilde \pi_i(b^{c,*})-t_i) \ge 0$ by Assumption \ref{as_unica}.
\end{proof}

\REC*

\begin{proof}
	It is easy to see that for each interval $(\alpha_j,\beta_j]$ returned by  $\textsc{Rec}((\alpha_j,\beta_j],p,\eta)$ it holds $P_{b^e \sim \gamma^e}(\exists i \in N_e: b^e_i \in (\alpha_j,\beta_j])\le p$ or $\beta_j-\alpha_j\le \eta$ and that the union of the returned intervals is $(0,1]$.
	Then, we prove that $k^*\le \frac{2 n_e}{p} \log{\mleft( \frac{1}{\eta}\mright)}$.
	Suppose by contradiction that Algorithm~\ref{alg} returns $k^* > \frac{2 n_e}{p}\log{\mleft( \frac{1}{\eta}\mright)}$ intervals $\{(\alpha_j,\beta_j]\}_{j \in [k^*]}$.
	Each of these intervals has been generated by the recursive call of Algorithm~\ref{alg} $REC((\alpha_j,\beta_j],p,\eta)$ from an interval $I_h$ such that $P_{b^e \sim \gamma^e}(\exists i \in N_e: b^e_i \in I_h)> p$ and $(\alpha_j,\beta_j]=(\alpha_h,\frac{\alpha_h +\beta_h}{2}]$ or $(\alpha_j,\beta_j]=(\frac{\alpha_h +\beta_h}{2}, \beta_h]$.
	We call $\mathcal{I}=\cup_{h \in [k^*]} \{I_h\}$ the set of all such intervals $I_h$. 
	If an interval does not respect the condition at Line~\ref{line}, Algorithm~\ref{alg} performs two recursive calls and, therefore, that interval can be seen as an internal node of a binary tree. 
	Vice versa, each interval satisfying such condition is a leaf node. 
	In a binary tree the number of internal nodes is at least half the number of leaf nodes, hence the cardinality of $\mathcal{I}$ is at least $C = \frac{k^*}{2} > \frac{n_e}{p}\log{\mleft( \frac{1}{\eta}\mright)}$.
	Moreover, at least $\frac{C}{\log{\mleft( \frac{1}{\eta}\mright)}} > \frac{\frac{n_e}{p}\log{\mleft( \frac{1}{\eta}\mright)}}{\log{\mleft( \frac{1}{\eta}\mright)}}= \frac{n_e}{p}$ intervals $I_h \in \mathcal{I}$ are disjoint. 
	We denote by $\mathcal{I}_d$ the set of such disjoint intervals.
	This results follows from the tree representation introduced above. 
	Each interval, corresponding to a node $N$, is disjoint from another interval, corresponding to a node $N'$, if $N$ is neither a parent nor a child of $N'$.
	We refer to $N$ and $N'$ as disjoint nodes.
	Notice that all the nodes in the same level are disjoint.
	The number of such nodes can be bounded by $\frac{C}{h}$, where $h$ is the depth of the tree.
	In particular, the $C$ internal nodes are partitioned over $h-1$ levels of the tree. Thus, there exists at least a level with $\frac{C}{h-1}$ nodes.
	The result follows from the fact that the maximum depth is at most $\log{\frac{1}{\eta}}$.
	This implies that $\sum_{I \in \mathcal{I}_d} P_{b^e \sim \gamma^e}( \exists i \in N_e: b^e_i \in I) > \frac{n_e}{p} p = n_e$. We reach a contradiction since $\sum_{I \in \mathcal{I}^*} P_{b^e \sim \gamma^e}(\exists i \in N_e: b^e_i \in I) 
	\le n_e$ for each set $\mathcal{I}^*$ of disjoint intervals.
	
	We conclude the proof showing that the algorithm runs in polynomial time in $n_e$, $1/p$, and $log(1/\eta)$.
	Recall the tree representation described above: the number of recursive calls of Algorithm~\ref{alg} is equal to the number of nodes which are at most twice the number of leaf nodes in a binary tree. 
	Therefore the recursive calls of the algorithm are at most $2k^* \leq  \frac{4n_e}{p} \log{\mleft( \frac{1}{\eta}\mright)}$.
\end{proof}

\DiscretizedBid*
\begin{proof}
	Recall that $I^{p,\eta} \coloneqq \{(\alpha_j,\beta_j]\}_{j \in [k^*]}$ is the set of intervals returned by $\textsc{Rec}((0,1], p,\eta)$.
	Take a bid profile $b^c \in B^c$.
	Consider the bid profile $\hat b^c \in B^{c,p}$ such that, for each $i \in N_c$, $\hat b^c_i$ is the largest element in $\bigcup_{(\alpha,\beta] \in I^{p,\eta}} \{\alpha\}$ such that $\hat b^c_i \le b^c_i$.
	Then, we increase each bid $\hat{b}^c_i$ by $\tau (n_c-i)$, for all $i \in [n_c]$ in order to have the same ordering of the colluders' bids in $b^c$ and $\hat b^c$. 
	This step is equivalent to introducing a specific tie breaking rule.
	Now, we show that the two conditions stated in the lemma hold for $\hat b^c$.
	First, it is easy to see that since each colluder decreases his bid (ignoring the arbitrary small $\tau$), the payment decreases both in VCG and GSP auctions.
	Then, we show that the revenue of each colluder $i \in N_c$ decreases by a small amount. In particular, since with probability at least $1-p$ there is not an external bid in the interval $(\hat b^c, b^c)$ and the partial ordering of the colluders's bids does not change, with probability at least $1-p$ the colluder $i$ is assign to the same slot. Hence, his utility is at least $\tilde r_i( b^c)-p$.
	This concludes the proof.
\end{proof}

\Arbitrary*

\begin{proof}
	Let $ b^{c,*} \in \argmax_{b^c \in B^{c}} \sum_{i \in N_c} \tilde r_i( b^{c}) - \tilde \pi(b^c)$. 
	By Lemma \ref{lm:discretized}, there exists a discretized bid profile $\tilde b^c \in B^{c,p}$ such that:
	\begin{itemize}
		\item $ \tilde \pi_i(\tilde b^c) \le \tilde \pi_i( b^{c,*})  \ \forall i \in N_c$
		\item $ \tilde r_i(\tilde b^c) \ge \tilde r_i( b^{c,*}) -p \ \forall i \in N_c $.
	\end{itemize}
	Hence, $\sum_{i \in N_c} \tilde r_i( \hat b^{c}) - \tilde \pi_i(\hat b^{c}) \ge \sum_{i \in N_c} \tilde r_i( \tilde b^{c}) - \tilde \pi_i(\tilde b^{c})  \ge \sum_{i \in N_c} \tilde r_i( b^{c,*}) - \tilde \pi_i(b^{c,*}) - n_c p$.
	We consider the solution to Problem~\eqref{lp:problem1}
	composed of bidding strategy $\hat b^{c}$, with and $q_i=\tilde r_i(\hat b^{c})- t_i +p$.
	The solution has objective at least $OPT-n_c p$ and satisfies $p$-IC by construction.
	To conclude the proof, we show that the solution is IR.
	In particular, it holds $\sum_{i \in N_c} (q_i -\tilde \pi_i(\hat b^{c}))= \sum_{i \in N_c} \mleft(\tilde r_i(\hat b^{c}) -\tilde \pi_i(\hat b^{c}) - t_i \mright) +pn_c \ge \sum_{i \in N_c}  \mleft(\tilde r_i( b^{c,*}) -\tilde \pi_i( b^{c,*}) - t_i \mright) \ge  0$, where the last inequality follows from Assumption \ref{as_unica}.
\end{proof}

\Feasible*
\begin{proof}
	We show how to construct a solution that satisfies the two constraints.
	Take the optimal solution $\gamma$, $q$ to LP \eqref{lp:problem1} with LL. 
	Let $S:B^c\rightarrow B^{c,p}$ be the function that maps each bid profile $b\in B^c$ to the bid profile $b' \in B^{c,p}$ that satisfies Lemma \ref{lm:discretized}.
	We build a solution $\bar \gamma$, $\bar q$ to LP \eqref{lp:Limit} such that $\bar \gamma_{ b^c}=Pr_{ \tilde b^c\sim \gamma}( {b}^c = S( \tilde b^c) )$ for each $b^c \in B^{c,p}$. Moreover, we take $\bar q_i=q_i$ for each $i \in N_c$.
	By lemma \ref{lm:discretized}, we have that for each $b^c \in B^c$ and $i \in N_c$, $\tilde r_i(S(b^c))\ge r_i(b^c)-p$ and $\tilde \pi_i(S(b^c))\le  \pi_i(b^c)$.
	Hence, $\sum_{b^c \in B^{c,p}} \bar \gamma_{b^c} \sum_{i \in N_c} \tilde \pi_i(b^c)\le E_{b^c \sim \gamma} [ \sum_{i \in N_c} \tilde \pi_i(b^c) ] $ and $\sum_{b^c \in B^{c,p}} \bar \gamma_{b^c} \sum_{i \in N_c} \tilde r_i(b^c)\ge E_{b^c \sim \gamma} [ \sum_{i \in N_c} \tilde r_i(b^c) ]-p $. 
	Thus, the objective decreases of at most $p n_c$ while $\sum_{b^c \in B^{c,p}} \bar \gamma_{b^c} \tilde r_i(b^c)-\bar q_i \ge E_{b^c \sim \gamma} [ \tilde r_i(b^c)]-p -q_i\ge t_i-p$ for each $i \in N_c$ and Constraints \eqref{c_inc2} are satisfied.
	Finally, $\sum_{i \in N_c} \bar q_i = \sum_{i \in N_c}  q_i \ge E_{b^c \sim \gamma} [ \sum_{i \in N_c} \tilde \pi_i(b^c) ] \ge  \sum_{b^c \in B^{c,p}} \bar \gamma_{b^c} \sum_{i \in N_c} \tilde \pi_i(b^c)$.
	This concludes the proof.
\end{proof}

\ExistPloyAlgorithm*
\begin{proof}
	By Lemma \ref{lm:feasible}, to provide the desired guarantees it is enough to provide a solution to LP \eqref{lp:Limit} with $p=\epsilon/ n_c$.
	By strong duality, it is sufficient to solve the dual LP \eqref{lp:dual}.
	The algorithm employs the ellipsoid method to solve the dual. To do so, it needs a polynomial-time separation oracle that given an assignment $(y,x,z)$ to the variables returns a violated constraint (if any). 
	Since there are only polynomially-many Constraints \eqref{dual:3}, we can check if one of these constraints is violated in polynomial time.
	Moreover, we can find in polynomial time if the exists a violated Constraint \eqref{dual:2} solving $\max_{b^c \in B^{c,p}} \sum_{i \in N_c} [(1-y_i)\tilde{r}_i(b^c)+(x-1)\tilde\pi_i(b^c)]$. This is an instance of the weighted utility problem in Equation \eqref{obj_wup} with $\hat{y_i} =1-y_i$ for each $i \in N_c$ and $\hat x=x-1$. If the value is grater than $z$, we return the constraint relative to the solution of Equation \eqref{obj_wup}. Otherwise, all the constraints \eqref{dual:2} are satisfied. Finally, \eqref{obj_wup} can be solved in polynomial time by Theorem \ref{th_poly}.
	This concludes the proof.
\end{proof}

\end{document}